\documentclass[11pt,twoside]{article}
\usepackage{asp2014}

\aspSuppressVolSlug
\resetcounters
\begin{document}

\title{Deep unsupervised domain adaptation applied to the Cherenkov Telescope Array Large-Sized Telescope}

\author{Michael Dellaiera,$^{1,2}$ Cyann Plard,$^1$ Thomas Vuillaume,$^1$ Alexandre Benoit,$^2$ and Sami Caroff$^1$}
\affil{$^1$Univ. Savoie Mont Blanc, CNRS, LAPP, Annecy, France; \email{lastname@lapp.in2p3.fr}}
\affil{$^2$Univ. Savoie Mont Blanc, LISTIC, Annecy, France}

\paperauthor{Michael Dellaiera}{dellaiera@lapp.in2p3.fr}{0000-0002-5221-0240}{Univ. Savoie Mont Blanc, CNRS, LAPP}{IN2P3}{Annecy}{}{74000}{France}
\paperauthor{Thomas Vuillaume}{thomas.vuillaume@lapp.in2p3.fr}{0000-0002-5686-2078}{Univ. Savoie Mont Blanc, CNRS, LAPP}{IN2P3}{Annecy}{}{74000}{France}
\paperauthor{Alexandre Benoit}{alexandre.benoit@univ-smb.fr}{0000-0002-0627-4948}{Univ. Savoie Mont Blanc, LISTIC}{}{Annecy}{}{74000}{France}
\paperauthor{Cyann Plard}{plard@lapp.in2p3.fr}{0000-0002-4061-3800}{Univ. Savoie Mont Blanc, CNRS, LAPP}{IN2P3}{Annecy}{}{74000}{France}
\paperauthor{Sami Caroff}{caroff@lapp.in2p3.fr}{0000-0002-1103-130X}{Univ. Savoie Mont Blanc, CNRS, LAPP}{IN2P3}{Annecy}{}{74000}{France}



\begin{abstract}
The Cherenkov Telescope Array Observatory (CTAO) is the next generation of observatories employing the imaging air Cherenkov technique for the study of very high energy gamma rays. The deployment of deep learning methods for the reconstruction of physical attributes of incident particles has evinced promising outcomes when conducted on simulations. However, the transition of this approach to observational data is accompanied by challenges, as deep learning-based models are susceptible to domain shifts. In this paper, we integrate domain adaptation in the physics-based context of the CTAO and shed light on the gain in performance that these techniques bring using LST-1 real acquisitions.
\end{abstract}



\section{Introduction}
The Cherenkov Telescope Array (CTA) is the next generation of gamma-ray observatory and aims to improve the sensitivity by 5-10 times of the current ground-based instruments, and covers an energy range from 20 GeV to over 300 TeV. The first Large-Sized Telescope (LST-1) prototype is operational, successfully detecting known sources. The observation of gamma rays is an indirect process and uses the atmosphere as a calorimeter to catch Cherenkov radiation from particle interactions. The telescope acquisitions are processed to determine the incident particle physical parameters, but current methodologies fail to tackle the domain shift between the labelled training simulations and the test real data \citep{zhao2020}. Our work applies physics-guided deep learning to gamma-ray astronomy, incorporating domain adaptation into a sophisticated deep-learning architecture for IACT event reconstruction, as proposed in \citep{jacquemont2021}, focusing on multitask balancing and performance evaluation.

\section{Related work}
Imaging Atmospheric Cherenkov Telescopes (IACTs) analysis traditionally uses classical machine learning algorithms with data-engineered features. The Hillas method, developed by \citep{hillas1985}, is foundational in this domain, extracting image parameters used with random forests \citep{ohm2009} to predict particle characteristics, hereafter noted as \textit{Hillas+RF}. This method, however, becomes less effective at lower energy levels due to challenges in reconstructing faint images and information loss from noise removal procedures.

In contrast, recent advances involve full-event reconstruction \citep{miener2021} using deep learning models such as the $\gamma$-PhysNet, introduced by \citep{visapp2021}. This model, differing from the Hillas+RF mono-task approach, reconstructs all parameters simultaneously using a ResNet encoder with attention mechanisms and a multi-task framework. While the $\gamma$-PhysNet has shown improvements, particularly in detecting more gamma rays than standard analysis, challenges remain when applied to real telescope data. These include biases in spatial reconstruction and discrepancies between simulations and real-world data, as investigated in \citep{jacquemont2021} and \citep{vuillaume2021}, with encouraging results observed in controlled settings.

\section{Present work}
In this work, we implemented two domain adaptation methods that are DANN \citep{ganin2016} and DeepCORAL \citep{tzeng2014} to tackle deep unsupervised domain adaptation with multi-task learning in the context of real data analysis. They constitute a simple yet relevant baseline to start building more sophisticated algorithms in the future. On the one hand, DANN is a state-of-the-art domain adaptation model that can extend almost any feed-forward classification or regression model by the addition of a domain classifier in parallel to the reconstruction tasks. The domain classifier can either be trained to use the real data and simulated gammas/protons, or only using real data and source protons, in order to take into account the label shifts. In the latter, the network is referred to as DANN conditional. On the other hand, DeepCORAL aims to minimize the distance of the second-order statistics of the source and target distributions. The correlation alignment quantifies the misalignment or distance between the covariance matrices. In both cases, the real data sampling is as follows: real data are stored in files of thousands of images sorted by time. We uniformly sample a selection of files and extract all the images, corresponding to $5\%$ of the total amount.

\section{Results}
The crab nebula is a standard and bright source used by many gamma-ray observatories due to its high flux and expected flux stability. In this work, we use a dataset of LST-1 observations \citep{abe2023} composed of four consecutive observation runs with their observation conditions described in Table \ref{tab:crab_conditions}. This dataset is divided into two samples of two runs, respectively with and without moonlight, given the fact that the moonset happens between runs $6893$ and $6894$. Light pollution is defined using the mean number of photons observed per pixels for images taken at random time (uncontaminated by cosmic ray showers). For each of the observation sample, we compare the performance of $\gamma$-PhysNet with the data reconstruction pipeline of LST-1 implemented in \texttt{lstchain v0.9.6} \citep{lopez2021}, by estimating the significance ($\sigma$) \citep{lima1983} of the source in each sample.

\renewcommand{\arraystretch}{1.5} 
\begin{table}
    \centering
    \begin{tabular}{c|cc|cc}
         & 6892 & 6893 & 6894 & 6895 \\ 
        \hline
        Zenith angle (deg) & $16.1$ & $20.3$ & $27.9$ & $32.4$ \\
        Light pollution & $1.94$ pe & $1.81$ pe & $1.64$ pe & $1.60$ pe \\
        
    \end{tabular}
    \caption{Sample of crab data used as consecutive observation runs with their corresponding observation conditions.}
    \label{tab:crab_conditions}
\end{table}

We use the runs 6892 and 6894 to optimize the gammaness cut of the analysis assuming that the observation conditions are similar to the next run. Gammaness optimization is done by selecting the gammaness maximizing the significance of the Crab Nebula for those runs. We then extract the Crab Nebula significance by using the second run with the optimized gammaness selection, respectively 6893 and 6895. We also checked that the background levels matched between experiments, ensuring that the significance was not artificially increased for a specific analysis. The significance of this known source is used as an estimator of the performance of the analysis chains. The results are presented in Figure \ref{fig:results} for data recorded with or without background moonlight.

\articlefigure[width=\textwidth, trim={3cm 0.9cm 3cm 1cm}, clip]{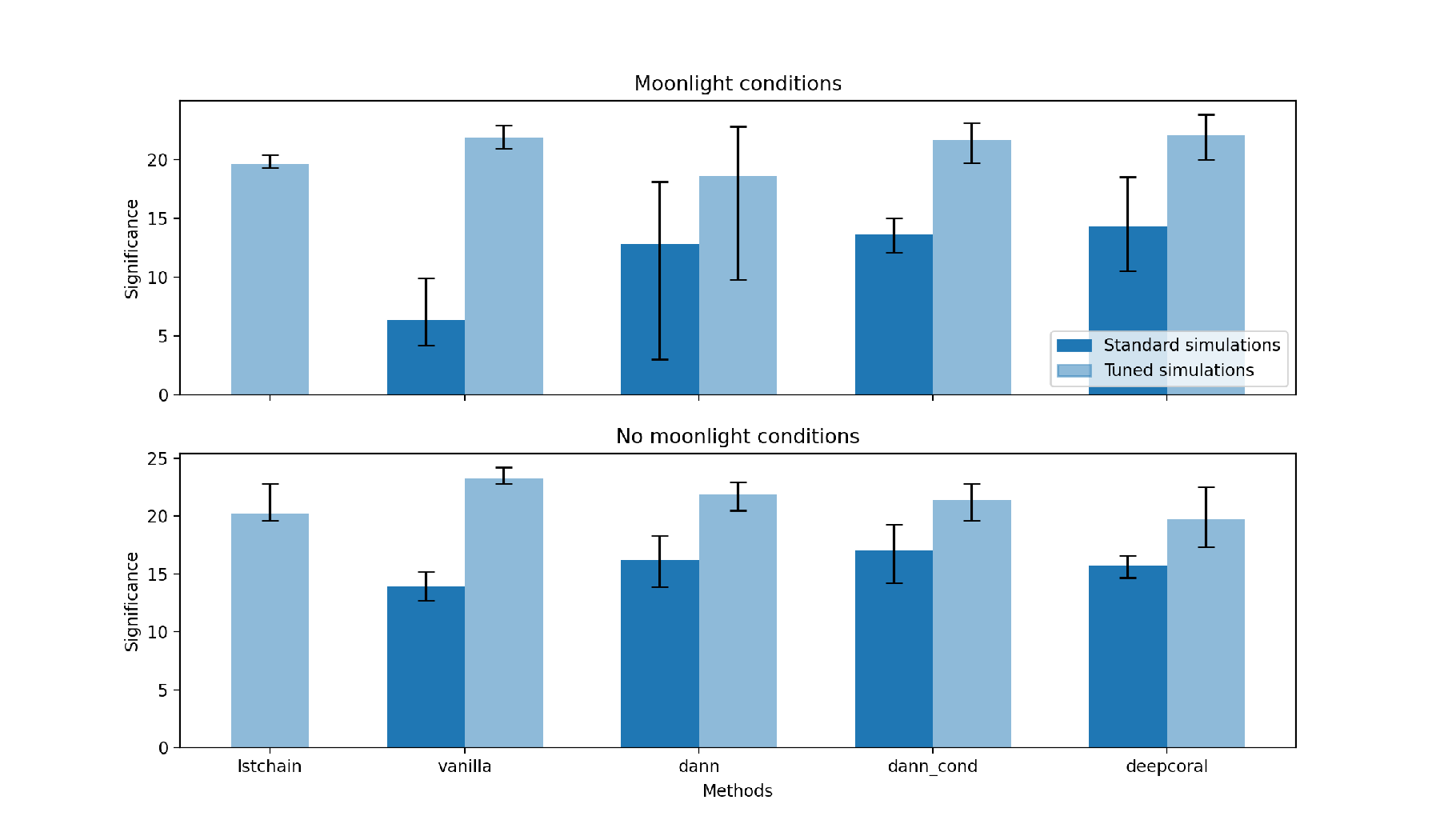}{fig:results}{Comparison of the different reconstruction methods trained with standard (not tuned) simulations or simulations with background levels tuned to the NSB level measured in the data analysed with moonlight background (top) and without moonlight background (bottom)}

\section{Conclusion}
As in previous works \citep{vuillaume2021, parsons2022}, we note that CNNs are very sensitive to NSB level, as showed by the difference of significance with and without the addition of noise in the source dataset. We demonstrate that domain adaptation methods can partially recover the loss of significance by matching the source and target distributions. Notably, the recovery of significance is more important in harsher conditions (with moonlight). However, tuning the training simulations to match the background level of the test dataset leads to the best results, with or without domain adaptation. Even more interestingly, even with tuned simulations, adding domain adaptation methods does not lead to an increase of significance compared to the Vanilla version, but to an increase in the variance as shown by the error bars in Figure \ref{fig:results}.
This shows two things: 1. The NSB level is most probably the most significant source of discrepancy between the data and 2. Domain adaptation methods do not seem to be able to compensate for other sources of discrepancies (stars in the field of view, atmospheric and weather conditions, reflectivity of the optics, calibration systematic uncertainty...) between simulations and these specific observational data.  On the contrary, the domain task appears to disrupt the primary reconstruction tasks, introducing greater variance in the reconstruction process, similar to results obtained on simulated data by \citep{dellaiera2023}. Domain adaptation may be interesting for most extreme cases of discrepancy, this will be explored in future studies.

\acknowledgements We gratefully acknowledge financial support from the agencies and organizations listed here\footnote{\\\url{https://purl.org/gammalearn/acknowledgements}} and from the French Programme investissements avenir through the Enigmass Labex as well as feedback from CTA members.

\bibliography{P4-06} 

\end{document}